# The Atomic-scale Growth of Large-Area Monolayer Graphene on Single-Crystal Copper Substrates

L. Zhao[1], K. T. Rim[2], H. Zhou[1], R. He[1], T. F. Heinz[1,2], A. Pinczuk[1,3], G. W. Flynn[4], A. N. Pasupathy[1]

[1]Department of Physics, Columbia University, New York NY 10027.
[2]Department of Electrical Engineering, Columbia University, New York NY 10027.
[3]Department of Applied Physics and Applied Mathematics, Columbia University, New York NY 10027.
[4]Department of Chemistry, Columbia University, New York NY 10027

We study the growth and microscopic structure of large-area graphene monolayers, grown on copper single crystals by chemical vapor deposition (CVD) in ultra-high vacuum (UHV). Using atomic-resolution scanning tunneling microscopy (STM), we find that graphene grows primarily in registry with the underlying copper lattice for both Cu(111) and Cu(100). The graphene has a hexagonal superstructure on Cu(111) with a significant electronic component, whereas it has a linear superstructure on Cu(100). The film quality is limited by grain boundaries, and the best growth is obtained on the Cu(111) surface.

The successful growth of large-area, few-layer graphene films[1-4] has the potential to revolutionize applications of graphene in electronic and mechanical devices. Recently, CVD growth has been used to realize such films on metal surfaces. In this technique, hydrocarbon gas is passed at millitorr pressures and high temperatures (800-1100°C) over metal substrates such as Cu[2, 5], Ni[6-8], Ru[9-12] and Ir[13-16], resulting in the formation of graphitic films on the surface. Cu is an especially important substrate due to its low cost, and the fact that graphene films grown on Cu foils are predominantly one monolayer thick[2]. In order to improve the quality of these CVD grown films, it is necessary to characterize their microscopic structure and determine the factors that limit film quality. Scanning tunneling microscopy (STM) is an atomic-resolution tool that has previously been used to probe the microscopic structure of epitaxial graphene films on silicon carbide[17] and metal substrates[13, 18-20]. Clean STM experiments can be performed on graphene films that have been grown in-situ on single-crystal surfaces, thus avoiding exposure of the sample to ambient conditions [9, 13, 17-18, 20]. We describe here the first experiments probing the large-area growth of graphene on single crystal Cu substrates using in-situ STM.

Our experiments were carried out in a UHV chamber with low ($10^{-10}$ torr) base pressure, using a variable temperature STM with sub-picometer resolution. Graphene films were grown on single crystal Cu(111) and Cu(100) substrates that are several mm in diameter and ~ 1 mm thick. Shown in Fig. 1(a) is a typical pristine Cu(111) surface. The surface shows well ordered terraces[21] that are ~ 100 nm wide. In previous in-situ experiments on other single crystal surfaces, graphene was grown by passing ~ 100 L of hydrocarbon over heated metal substrates[18, 20, 22]. Accordingly, we first attempted to grow graphene on Cu (111) by heating the crystal up to 875°C in ethylene at pressures in the $10^{-5}$ torr range for up to 20 minutes (up to $10^4$ L). This was followed by annealing the sample at 800°C for 15 minutes at $10^{-9}$ torr and cooling down to room temperature. We found that these conditions are insufficient for growing large-area graphene on the surface, indicating that the catalytic efficiency of copper is much lower than that of Ni, Ru, Ir and Co. We did, however, find evidence of carbon incorporation into the copper crystal - the well-ordered terraces on the pristine Cu surface disappear and are replaced by a rough topology with several islands and valleys as shown in Fig. 1(b). The step heights for all the islands and valleys in the figure correspond to the Cu (111) step height, and the atomic structure on the islands also shows only the hexagonal lattice of copper. Sequential scans taken over the surface show that these copper islands and depressions can diffuse across the surface on relatively short timescales, as shown in Fig. 1(c)-(d), and can even merge or break up. When these rough surfaces are annealed at high temperatures (800°C) for extended periods of time (several hours), the islands and depressions combine to form well-ordered terraces, as shown in Fig. 1(e). Occasionally (probability <1%), after annealing the surfaces exhibit nanoscale graphene islands that have different step heights (Fig. 1(f)) from that of pristine copper. An atomic resolution image taken on the island shows the graphene honeycomb structure (inset, Fig. 1(e)). Interestingly, we find that the angle between the graphene lattice and the edge of this hexagonal island is 0°, implying that this graphene fragment has a zigzag structured edge. This growth recipe, therefore, provides a way of producing nanoscale islands of graphene that can be used in experiments designed to study the structure, electronic properties, and reactivity of graphene edges.

In order to prepare large area samples of graphene, we increased the ethylene exposure of the hot Cu surface. Accordingly, the clean Cu(111) crystal was exposed at 900°C to 1 mtorr of ethylene for 5 minutes (3 × $10^5$ L), which was then followed by annealing at 800°C for 15 minutes. The results of this procedure are shown in Fig 2(a). Over large areas (inset, Fig 2(a)), atomic terraces are observed just as for the clean copper surface. A close look at a terrace (Fig 2(a)) shows a hexagonal superstructure with a periodicity of 60+/-5 Å and a peak to trough height of 0.35±0.1 Å. Superstructures such as this are observed over the entire single crystal. When the surface is scanned with high resolution (Fig 2(b) and inset), the graphene honeycomb lattice is clearly visible. The sample was removed from the STM chamber and Raman spectroscopy was performed at various locations on the crystal. A typical spectrum is shown in Fig 2(c). We observe a sharp (width ~30 $cm^{-1}$), single Lorentz profile 2D peak at ~2700 $cm^{-1}$, and a ratio of ~0.3 for the G to 2D peaks. These are good indicators of the presence of high-quality monolayer graphene. We also observe a clear D peak in the spectrum indicating significant disorder. Overall, the quality of the graphene film is comparable to other CVD-grown graphene[3], but is inferior to the best exfoliated samples[23]. We discuss some possible reasons for this difference below.

The superstructure observed in Fig. 2(a) can be explained in a simple way as the "beating" formed by overlaying the graphene lattice on the copper lattice. Such superstructures can be easily visualized by placing the graphene honeycomb lattice over the copper hexagonal lattice as shown in Fig. 2(d). Different angles between the lattices create hexagonal superstructures with different periodicities, and the wavelength of the superstructure can in turn be used to determine this angle. If $\vec{k_c}$, $\vec{k_{Cu}}$ and $\vec{k_{Sup}}$ are the reciprocal space vectors corresponding to the graphene lattice, the Cu(111) lattice, and the superstructure lattice, respectively, then $\vec{k_{Sup}} = \vec{k_{Cu}} - \vec{k_c}$. In the case of Fig. 2(a), $k_{Sup}$ =(1/6.0) nm$^{-1}$ while $k_c=1/a_c$=(1/0.245) nm$^{-1}$ and $k_{Cu}=1/a_{Cu}$=(1/0.256) nm$^{-1}$. From these values, we find that the angle $\theta$ between the graphene lattice and underlying copper lattice is $\theta = 0^o$ for the domain in fig. 2(a). We find that the vast majority of the sample areas show this registry.

The electronic properties of the graphene and underlying copper can be probed by performing STM measurements at different tip-sample bias voltages. Shown in Fig. 3(a)-(c) are a sequence of STM topographs taken over the same area of the surface at various bias voltages. Three prominent features are seen in these topographs – point-like defects, electronic "rings" around the point defects and the hexagonal superstructure. The point-like defects are present at all bias voltages and have a depth of 0.4 Å. These defects look identical to point defects that are observed in pristine Cu (111) surfaces[21]. Since the overlying graphene lattice structure is observed to be locally unperturbed over these defects (not shown), these defects have their origin in the underlying copper lattice and not in the graphene layer. The second feature observed in the graphene-copper surface is the presence of "rings" around the copper point defects at bias voltages greater than -500 mV. These rings are similar to the scattering interference patterns seen on pure Cu(111) surfaces[21]. The wavelength and dispersion of these rings (Fig. 3(d)) are consistent with the Cu(111) surface state. A detailed study of these scattering patterns (currently in progress) can reveal the interaction between the copper surface state and the graphene. The third feature seen prominently in Fig. 3 is the hexagonal superstructure previously described. We note that while the wavelength of the superstructure is not energy dependent, the intensity of the pattern is strongly dependent on the bias voltage and becomes very weak below -500 mV, much like the scattering interference patterns. This implies that the superstructure observed in STM has a significant electronic component.

While a majority of the surface is covered by graphene oriented along the copper surface lattice direction, grain boundaries in the graphene are occasionally observed. Understanding the microscopic causes of such granularity is important for improving the quality of CVD-grown graphene films. Shown in Fig. 4(a)-(b) are two cases where grain boundaries are seen in STM topographs. Fig. 4(a) shows an image of an area of the sample which has a large number of nm-scale impurities. The superstructure orientation and wavelength are clearly different across the impurities, showing that graphene grain boundaries are stabilized by the presence of such large-scale impurities. In Fig. 4(b), we see that a graphene grain boundary exists at the edge of an atomic terrace of Cu(111). This kind of grain boundary is very rarely seen on the surface, and most graphene films are continuous across step edges as shown in Fig. 4(c). This type of grain boundary can set an ultimate limit on the quality of graphene films produced by CVD.

In several previous experiments[24-25] graphene has been reported to be impervious to many gases and chemicals and, thus, provides an atomically-thin protective coating for the surface to which it adheres. We have probed this property of graphene on the Cu (111) surface by exposing the graphene-covered surface to ambient atmosphere for 1 month and then imaging the surface without performing any sputter-anneal cleaning cycles. Under normal circumstances the reactive surface of copper would be completely oxidized, but we find (Fig. 4(d)) that the superstructure of graphene on copper can still be easily imaged, indicating the potential use of this substrate in studies of molecules and nanocrystals that are dispersed from solution.

While the experiments described above were carried out on Cu(111) surfaces, current CVD growth processes use copper foils[26] or copper thin films[27] which are polycrystalline. As a result, it is important to understand the growth of graphene on different copper crystal surfaces. We therefore performed graphene growth experiments on a Cu(100) single crystal, whose surface square lattice is very different from the graphene honeycomb lattice. The growth conditions were the same as those used to produce large-area graphene on the Cu(111) surface. We find that while most of the Cu(100) surface is covered with graphene (Fig. 5(a)), the micron-scale structure of the graphene shows grain boundaries as well as nanoscale valleys reaching down to the copper surface (inset 1, Fig. 5(a)). As a result, the graphene film has much poorer microstructure quality on Cu(100).

The graphene layer itself (Fig. 5(a)) shows the presence of a linear superstructure with a periodicity of 11 Å. Such a linear superstructure can be explained by overlaying the graphene lattice at an angle of ~0° with respect to the copper square lattice (Fig. 5(b)). Since there are two possible orientations in which the graphene lattice can be overlaid at 0° with respect to a copper lattice direction (rotated by 90° from each other), both orientations of graphene might be expected to exist on the surface with roughly equal probability. Consistent with this, Low Energy Electron Diffraction (LEED) patterns of the graphene-covered surface (inset 2, Fig. 5(a)) show two hexagons rotated by 90 degrees with respect to each other (the LEED spots are broad enough that the copper and graphene lattices are not distinguishable), as well as superstructure spots corresponding to the overlay of the graphene on the copper lattice. Besides the twelve strong Bragg spots, there is a faint ring at the same radius, which implies that there are also graphene islands randomly oriented with respect to the Cu(100) crystal lattice. Indeed, we observe that some areas of the sample display a 2-dimensional superstructure indicating the presence of graphene grains that are not aligned with the copper lattice. From all these observations we see that the graphene film properties are much poorer on the Cu(100) surface when compared to the Cu(111) surface. This is also confirmed through Raman spectroscopy measurements of the grown film (Fig 5(d)). The 2D peak was broadened (line width ~40cm$^{-1}$), the ratio of G to 2D peak increased up to 0.6, and, most importantly, the D peak is strongly enhanced on the Cu(100) surface when compared with the Cu(111) surface. These results indicate that the crystallographic orientation of the copper grains is of fundamental importance in determining the quality of graphene films produced, and attempts should be made to improve the copper substrate quality in order to achieve better graphene growth.

We thank D. Eom and W. Bang for assistance in STM measurements. This work was funded by AFOSR under grant no. FA9550-10-1-0068 (A.N.P); by the DOE under grant nos. DE-FG02-88ER13937 (G.W.F), DE-FG02-07ER15842 (T.H.) and EFRC Award DE-SC0001085 (G.W.F., T.H., A.P., A.N.P.); by ONR under Graphene MURI (A.P.); by NSF under grant no. CHE-0641523 (A.P.); and by NYSTAR. Equipment support was provided by the NSF under grant CHE-07-01483 (G.W.F.).

**References**
[1] C. Berger, *et al.*, J. Phys. Chem. B **108**, 19912-19916 (2004).
[2] X. S. Li, *et al.*, Science **324**, 1312-1314 (2009).
[3] K. S. Kim, *et al.*, Nature **457**, 706-710 (2009).
[4] S. Bae, *et al.*, Nat. Nanotech. DOI 10.1038/nnano.2010.132.
[5] M. P. Levendorf, C. S. Ruiz-Vargas, S. Garg and J. Park, Nano Letters **9**, 4479-4483 (2009).
[6] C. H. Lui, *et al.*, Nature **462**, 339-341 (2009).
[7] A. Reina, *et al.*, Nano Res. **2**, 509-516 (2009).
[8] H. J. Park, J. Meyer, S. Roth and V. Skakalova, Carbon **48**, 1088-1094 (2010).
[9] D. Martoccia, *et al.*, Phys. Rev. Lett. **101**, 126102 (2008).
[10] P. W. Sutter, J.-I. Flege and E. A. Sutter, Nat. Mater. **7**, 406-411 (2008).
[11] P. Sutter, M. S. Hybertsen, J. T. Sadowski and E. Sutter, Nano Lett. **9**, 2654-2660 (2009).
[12] W. Moritz, *et al.*, Phys. Rev. Lett. **104**, 136102 (2010).
[13] J. Coraux, A. T. N`Diaye, C. Busse and T. Michely, Nano Lett. **8**, 565-570 (2008).
[14] P. Lacovig, *et al.*, Phys. Rev. Lett. **103**, 179904 (2009).
[15] R. Balog, *et al.*, Nat. Mater. **9**, 315-319 (2010).
[16] H. Chen, W. G. Zhu and Z. Y. Zhang, Phys. Rev. Lett. **104**, 186101 (2010).
[17] G. M. Rutter, *et al.*, Science **317**, 219-222 (2007).
[18] D. Martoccia, *et al.*, Physical Review Letters **101**, (2008).
[19] Y. Pan, *et al.*, Adv. Mater. **21**, 2777-+ (2009).
[20] D. Eom, *et al.*, Nano Letters **9**, 2844-2848 (2009).
[21] M. F. Crommie, C. P. Lutz and D. M. Eigler, Nature **363**, 524-527 (1993).
[22] P. W. Sutter, J.-I. Flege and E. A. Sutter, Nat Mater **7**, 406-411 (2008).
[23] K. I. Bolotin, *et al.*, Nature **462**, 196-199 (2009).
[24] J. S. Bunch, *et al.*, Nano Letters **8**, 2458-2462 (2008).
[25] E. Stolyarova, *et al.*, Nano Letters **9**, 332-337 (2008).
[26] X. Li, *et al.*, Science **324**, 1312-1314 (2009).
[27] A. Ismach, *et al.*, Nano Letters **10**, 1542-1548 (2010).

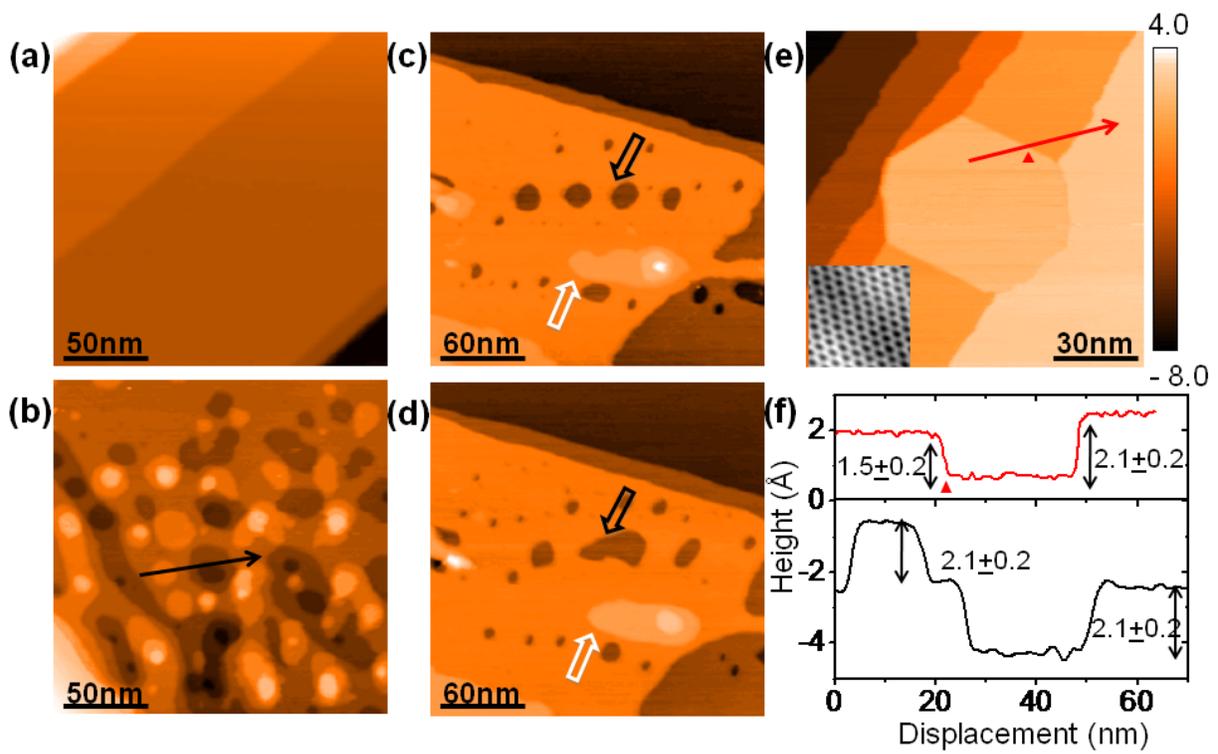

FIG 1 STM topographs (V=0.6V, I=1.0nA) of (a) a pristine Cu(111) surface and (b) after exposure to $10^4$ L of ethylene. (c),(d) Successive scans taken minutes apart on the ethylene-exposed copper surface, showing large-scale diffusion of copper on the surface. (e) nanoscale island of graphene after extensive annealing (f) STM line profiles from (b) (bottom) and from (e) (top)

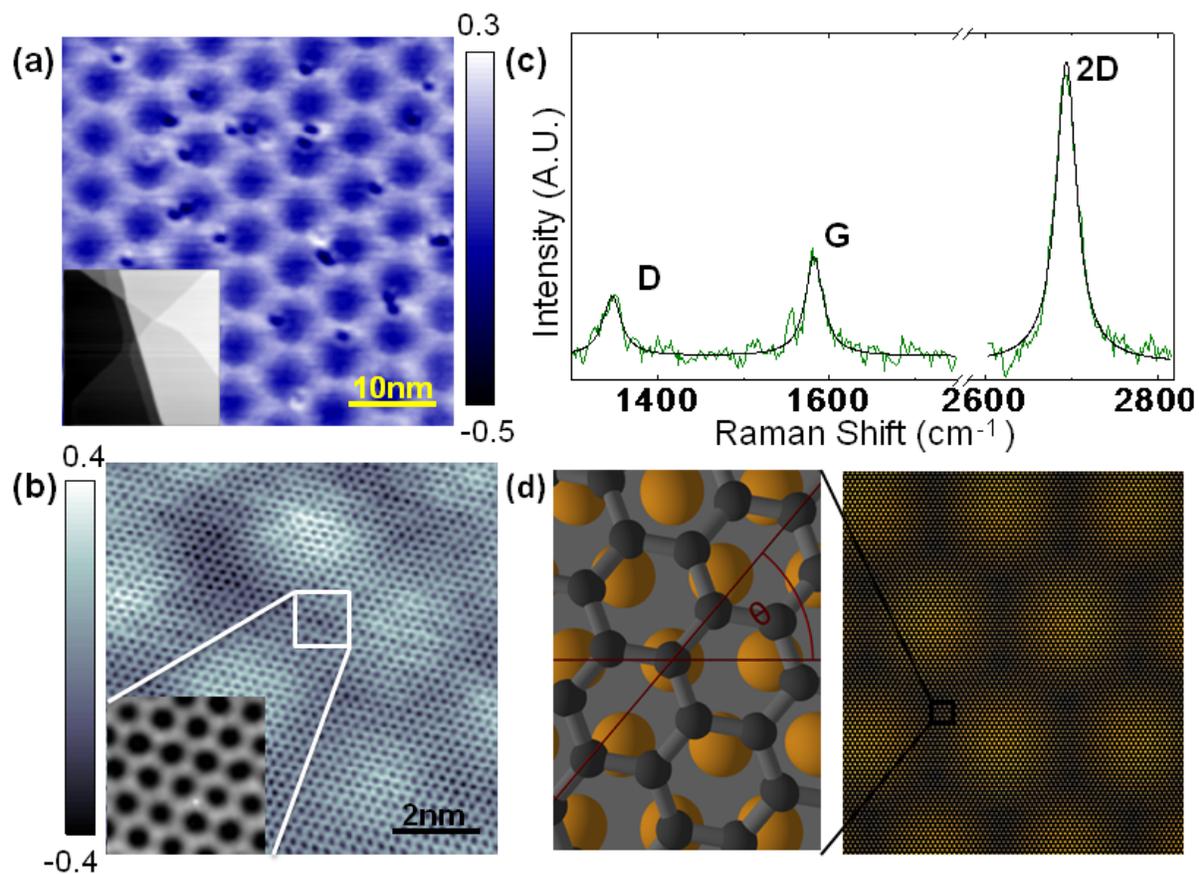

FIG 2 (a) Atomic terrace (V=0.6V, I=1.0 nA) after exposing clean Cu(111) to $3\times10^5$ L of ethylene at 900°C, showing a hexagonal superstructure across the surface, and point defects in the underlying Cu. (inset) $500\times500$ nm$^2$ topograph showing atomic terraces. (b) Higher resolution topograph showing the atomic scale structure of the graphene on the copper. (inset) honeycomb lattice of graphene. (c) Raman spectrum (after subtraction of Cu luminescence background) of CVD-grown graphene. (d) (left) schematic view of the overlay of graphene on Cu(111) (right) large-scale view of the resultant superstructure for the case θ=0.

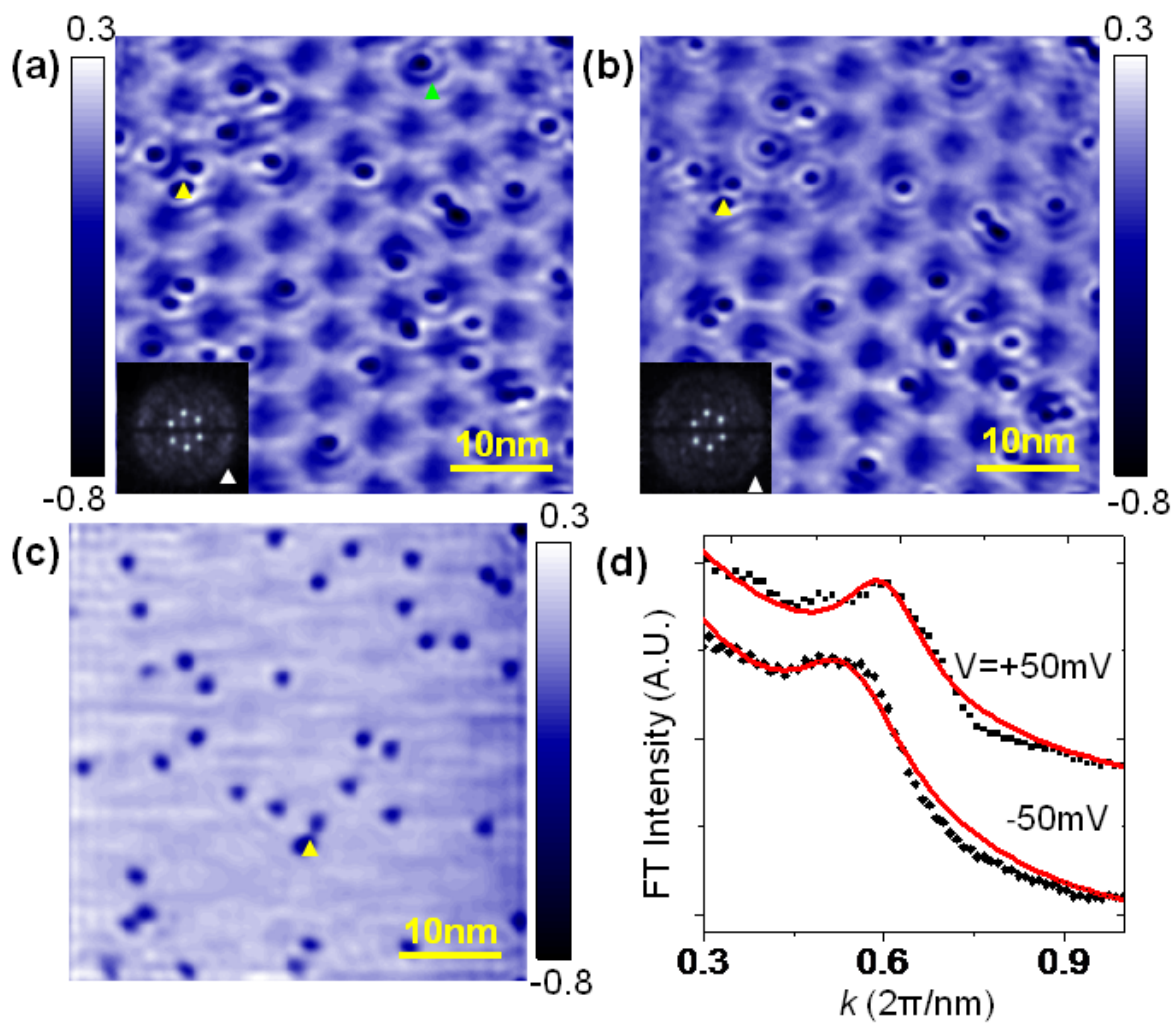

FIG 3 (a) Topograph taken at V= -50 mV, I=1.0 nA showing point defects (such as the yellow triangle), scattering interference rings around the point defects (such as the green triangle) and hexagonal superstructure. (b) Topograph of the same area at V=+50 mV, I=1.0 nA showing subtle differences in the ring structure from (a). Insets of (a) and (b) are FFT images, showing the hexagonal periodicity of the superstructure as well as a "disc" due to the scattering interference rings (white triangles). (c) Topograph at V= -500 mV showing the absence of the scattering patterns as well as the superstructure. (d) angle-averaged line profile of FFT's in (a) and (b), showing dispersing peaks at the scattering interference ring wavelength.

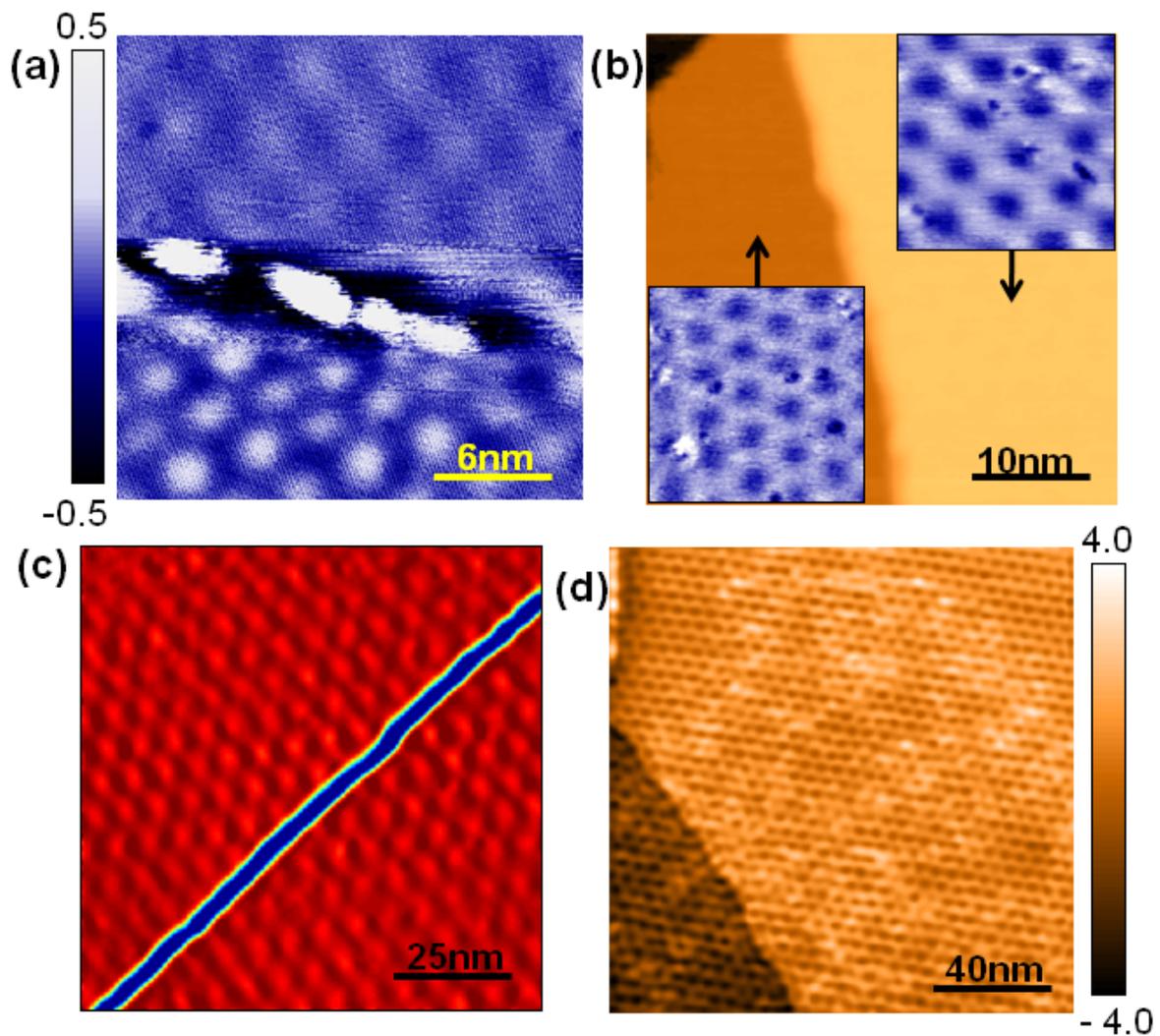

FIG 4 (a) Graphene grain boundary in Cu(111) caused by the presence of nanometer-scale impurities on the surface (b) Grain boundary at a copper step edge. The two insets show the graphene superstructure on either side of the step (c) More typical step edge in graphene where the superstructure is continuous across the edge of the step. The image is shown in derivative mode. (d) Topograph taken after exposing the graphene covered Cu(111) to air for 1 month. The superstructure is still clearly visible on the surface showing the protective properties of the graphene.

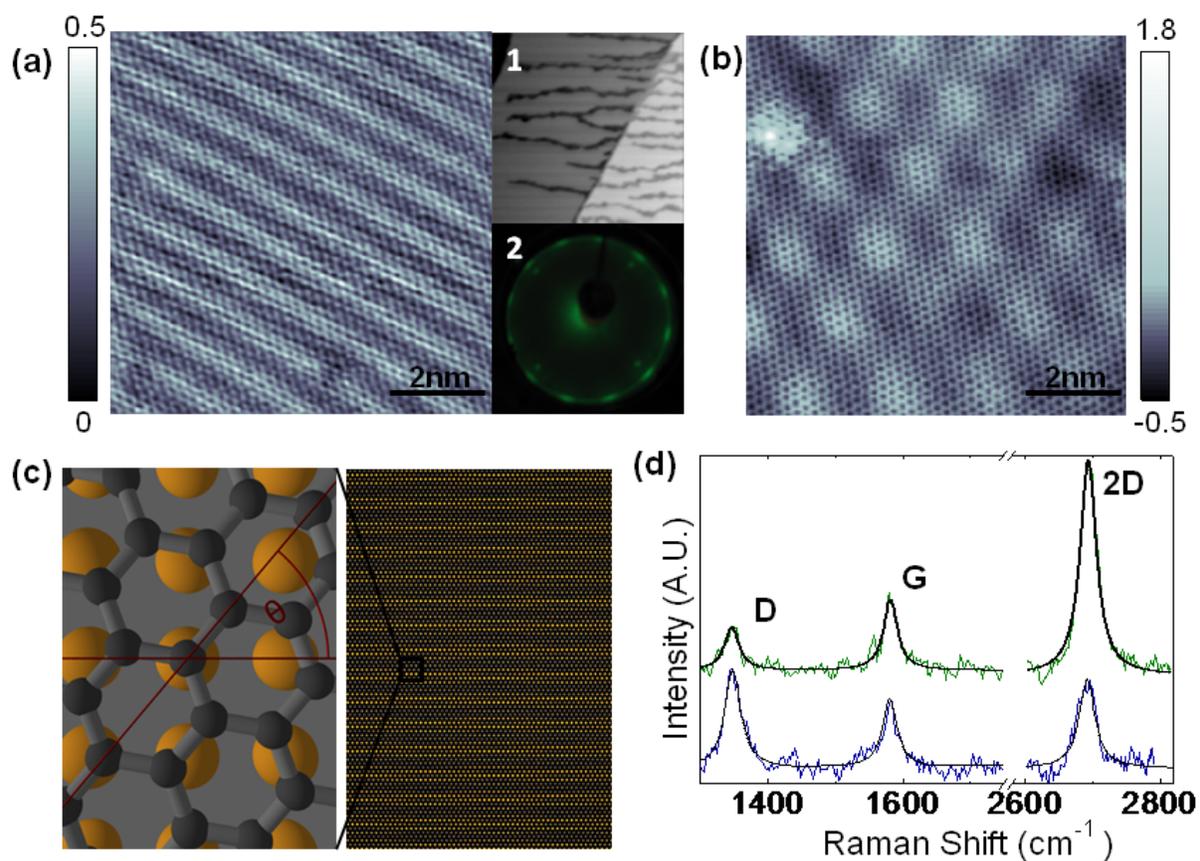

FIG 5 (a), Typical topograph (V=0.6 V, I=1.0 nA) of graphene grown on Cu(100) shows one-dimensional superstructure. Inset (1) is a 200×200 nm$^2$ topograph of the surface showing many nanoscale graphene edges on a terrace that reach down to the copper surface. Inset (2) is a LEED pattern showing two hexagons aligned with the two directions of the Cu(100) lattice, as well as a faint ring at the same radius. (b) Superstructure pattern in a different area of the sample. (c) (left) schematic view of the overlay of graphene on Cu(100) (right) large-scale view of the resultant one-dimensional superstructure for the case θ=0. (d) Raman spectra (Cu background subtracted) of graphene on Cu(100) (bottom) and Cu(111) (top).